\documentclass[9pt,twocolumn,twoside]{opticajnl}

\journal{opticajournal} 

\setboolean{shortarticle}{true}

\title{Photonic Mpemba effect}

\author[1,2,*]{Stefano Longhi}

\affil[1]{Dipartimento di Fisica, Politecnico di Milano, Piazza L. da Vinci 32, I-20133 Milano, Italy}
\affil[2]{IFISC (UIB-CSIC), Instituto de Fisica Interdisciplinar y Sistemas Complejos - Palma de Mallorca, Spain}

\affil[*]{stefano.longhi@polimi.it}

\begin{abstract}
The Mpemba effect is the counterintuitive phenomenon in statistical physics for which a far-from-equilibrium state can
relax toward equilibrium faster than a state closer to equilibrium. This effect has raised a great curiosity since long time and has been studied extensively in many classical and quantum systems. Here it is shown that the Mpemba effect can be observed in optics as well. Specifically, the process of light diffusion in finite-sized photonic lattices under  incoherent (dephasing) dynamics is considered. Rather surprisingly, it is shown that  certain highly-localized initial light distributions can diffuse faster than initial broadly delocalized distributions. The effect is illustrated by considering random walk of optical pulses in fiber-based temporal mesh lattices, which should provide an experimentally-accessible setup for the demonstration of the Mpemba effect in optics.
\end{abstract}

\setboolean{displaycopyright}{false} 

\begin{document}

\maketitle

{\em Introduction.}  The Mpemba effect (ME) \cite{S1,S2,S3} describes the counterintuitive physical phenomenon  where a far-from-equilibrium state can
relax to equilibrium faster than a state closer to equilibrium. For example, it is known since long time  that hot water can sometimes freeze faster than cold water \cite{S1,S2}.
The effect has been studied in many classical systems (see e.g. \cite{S3,S4,S5,S6} and references therein), and more recently extended to quantum systems \cite{S7,S8,S9,S10,S11,S12,S13,S14,S15,S16,S17}. Quantum ME  can be a non-thermal phenomenon and related to accelerated non-equilibrium relaxation in dissipative open systems \cite{S9} or can involve entanglement asymmetry in the dynamical restoration of symmetry after a symmetry-breaking quantum quench \cite{S11}.  Roughly speaking, in both classical and quantum markovian systems ME can be explained by the circumstance that certain far-from-equilibrium states display accelerated exponential decay in the relaxation process than states closer to equilibrium, avoiding to excite slow-decaying modes of the system \cite{S5,S9}. Although ME has been shown to arise in several area of physics, this phenomenon has been so far largely ignored by the photonic community. A main and interesting question then arises: can ME  be observed for light as well?\\
In this Letter we provide a simple example of ME in optics by considering random walks in photonic lattices under incoherent (dephasing) dynamics \cite{S18,S19,S19b,S20,S21,S22}. Remarkably, certain highly-localized initial light distributions can diffuse and reach uniform distribution in the lattice faster than initial highly-delocalized light distributions, a clear signature of the ME. The phenomenon is illustrated by considering incoherent photonic random walks based on optical pulse dynamics in fiber-based temporal mesh lattices.\\
\\
{\it Model and the Mpemba effect.}  Let us consider the quantum walk of bosonic particles, such a photons,  in a discrete one-dimensional lattice of coupled cavities or waveguides in the presence of dephasing effects [Fig.1(a)], which has been studied in several previous works \cite{S18,S19,S19b,S20,S21,S22,S23,S24,S25}. For a continuous-time walk \cite{S20,S21,S24}, the coherent dynamics  of the photon field is described by the tight-binding Hamiltonian
\begin{equation}
\hat{H}=\sum_{n=1}^{L-1} J  \left( \hat{a}^{\dag}_n \hat{a}_{n+1}+{\rm H.c.} \right)
\end{equation}
where $\hat{a}^{\dag}_n$, $\hat{a}_n$ are the bosonic creation and destruction operators at site $n$, $J$ is the coherent hopping rate (coupling constant) between adjacent sites, $L$ is the number of lattice sites, and open boundary conditions are assumed.
Dephasing effects for the bosonic field can be modeled using a quantum master equation in the Lindblad form for the density operator  $\hat{\rho}$ of the photon field (see e.g. \cite{S18,S21,S23,S24,S25,S26}), which reads 
\begin{equation}
\frac{d \hat{\rho}}{dt} = -i [ \hat{H}, \hat{\rho} ] + \gamma \sum_{n=1}^{L} \left( \hat{L}_n \hat{\rho} \hat{L}_n^{\dag}-\frac{1}{2} \left\{ \hat{L}_{n}^{\dag} \hat{L}_n, \hat{\rho} \right\}  \right) \equiv {\mathcal L} \hat{\rho} 
\end{equation}
 where $\hat{L}_n = \hat{a}^{\dag}_n \hat{a}_n$ is the dissipator describing pure dephasing at lattice site $n$, $\gamma>0$ is the dephasing rate, and $ {\mathcal L}$ is the Liouvillian superoperator. 
 This model conserves the total number of photons. In the single-photon sector of Hilbert space, after letting  $|n \rangle = \hat{a}_n^{\dag} | vac \rangle$ the evolution equations for the density matrix elements $\rho_{n,m}(t)= \langle n | \hat{\rho} |m \rangle$ read (see e.g. \cite{S23,S24,S26}) 
 \begin{eqnarray}
 \frac{d \rho_{n,m}}{dt} & = & i J (\rho_{n-1,m}+ \rho_{n+1,m}- \rho_{n,m+1}- \rho_{n,m-1})  \nonumber \\
 & - & \gamma (1- \delta_{n,m}) \rho_{n,m} \equiv \sum_{ \alpha, \beta=1}^L \mathcal{L}_{n,m; \alpha, \beta} \rho_{\alpha, \beta}
 \end{eqnarray}
 which should be supplemented with the open boundary conditions $\rho_{n,m}=0$ for $n,m > L$ and $n,m < 1$. Note that the Liouvillian superoperator $\mathcal{L}$ is described by the $L^2 \times L^2$ matrix entering in Eq.(3).
  \begin{figure}[h]
 \centering
    \includegraphics[width=0.42\textwidth]{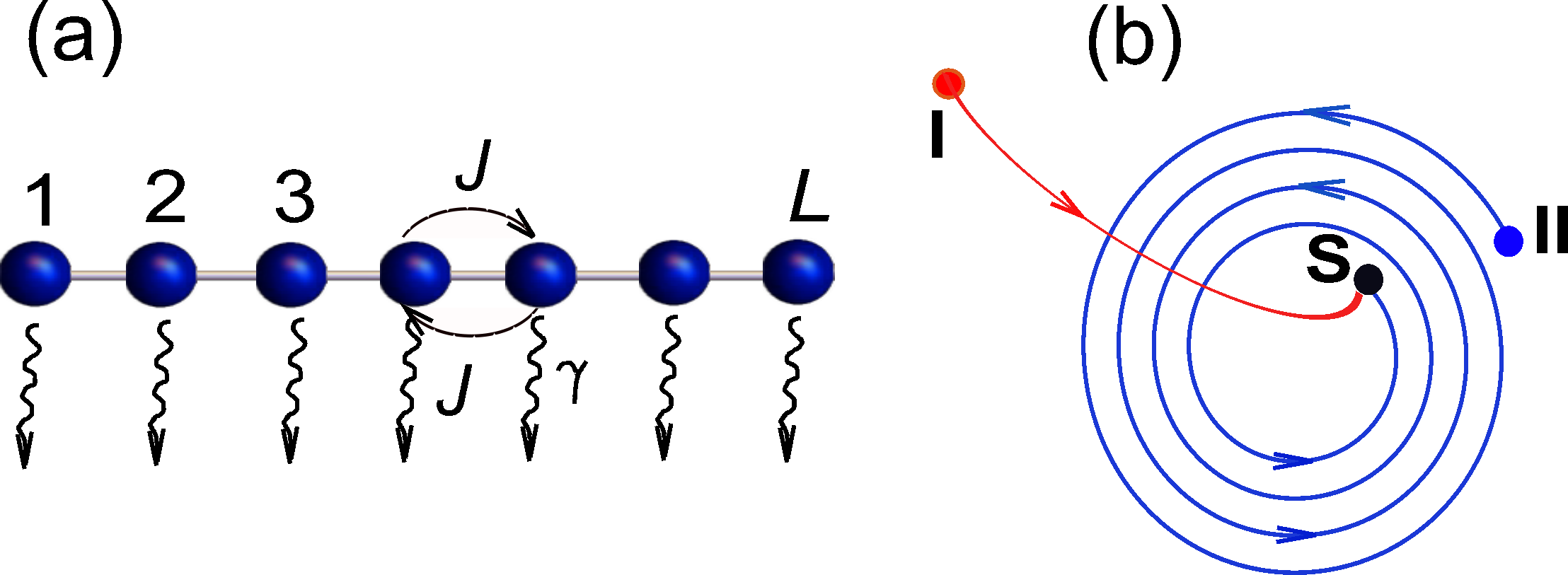}
   \caption{ \small (a) Schematic of photonic quantum walk on a one-dimensional lattice comprising $L$ sites (cavities or waveguides) with open boundary conditions. $J$ is the coherent hopping amplitude, $\gamma$ is the dephasing rate. In the strong dephasing regime $\gamma /J \gg 1$ the photon  undergoes a classical random walk on the lattice with an incoherent left/right hopping rate $J_{inc}=2J^2/ \gamma$. (b) Cartoon of the Mpemba effect. An initial far-from equilibrium state I can relax toward the equilibrium state S faster than a closer-to-equilibrium initial state II.}
\end{figure}
  The stationary state of the dynamics is given by the maximally-mixed state $\hat{\rho}^{(S)}=\rho_{n,m}^{(S)} =(1/L) \delta_{n,m}$, which is an eigenvector of $\mathcal{L}$ with zero eigenvalue. This stationary state basically corresponds to equal probability to find the photon in  either one of the $L$ lattice sites with vanishing coherences. All other eigenvalues $\lambda$ of $\mathcal{L}$ have a negative real part. The relaxation dynamics towards the stationary state is governed by such eigenvalues and corresponding eigenvectors excited by the initial state. In particular, the long-time dynamics is established  rather generally by the spectral gap of the Liouvillian, i.e. by the slowest decaying eigenstate of $\mathcal{L}$ (see e.g. \cite{S17}). The ME corresponds to the counterintuitive phenomenon for which an initial state I, very far from the stationary (equilibrium) state S, relaxes toward S faster than other initial state II closer to S; a cartoon of the ME is illustrated in Fig.1(b). To quantify the relaxation dynamics, let $\hat{\rho}_I(t)$ and $\hat{\rho}_{II}(t)$ be the solutions to Eq.(2), corresponding to two different initial conditions $\hat{\rho}_I(0)$ and $\hat{\rho}_{II}(0)$, and let $D_{\hat{\rho}} (t)$ be a measure of the 'distance' between the state $\hat{\rho}(t)$ and the stationary state $\hat{\rho}^{(S)}$. The definition of $D_{\hat{\rho}} (t)$ is not  unique, and different possibilities have been considered in previous works  (see e.g. \cite{S8,S9,S17}). Here we measure the distance $D_{\hat{\rho}} (t)$ in terms of the von Neumann entropy, namely we assume
  \begin{eqnarray}
  D_{\hat{\rho}} (t) & = & {\rm Tr} \left\{  \hat{\rho}(t) \log \hat{\rho}(t) \right\}-  {\rm Tr} \left\{ \hat{\rho}^{(S)} \log \hat{\rho}^{(S)} \right\} \nonumber \\
  & = & {\rm Tr} \left\{  \hat{\rho}(t) \log \hat{\rho}(t) \right\}+ \log L .
  \end{eqnarray}
  {\color{black} which is also equivalent to the distance function based on the Kullback-Leibler divergence \cite{S8}, given that $\hat{\rho}^{(S)}$ is diagonal. 
  Note that $D_{\hat{\rho}}(t) \geq 0$, vanishing when $\hat{\rho}(t)=\rho^{(S)}$. 
   Note also that, since decoherence induced by pure dephasing acts like an infinite-temperature bath for the photon field \cite{S27}, the distance $D$ based on the von Neumann entropy is equivalent to the measure function defined by the non-equilibrium free energy used in recent works \cite{S17}. }
  The ME arises whenever, for two assigned initial states such that  $D_{\hat{\rho}_I} (0)>D_{\hat{\rho}_{II}} (0)$, asymptotically for $t \rightarrow \infty$ one has $D_{\hat{\rho}_I} (t)<D_{\hat{\rho}_{II}} (t)$. Rather generally, the ME is observed when the initial state $\hat{\rho}_I (0)$ is far distant from the stationary state but does not excite the slowest decaying eigenmode of $\mathcal{L}$, resulting in a fast relaxation toward the stationary state, while $\hat{\rho}_{II} (0)$ is closer to the stationary state but has a non-vanishing overlapping with the slowest decaying mode of $\mathcal{L}$, resulting in a slower relaxation dynamics. {\color{black} For the quantum walk model, requirements for state I are satisfied by considering a highly-localized initial excitation of the lattice, which is dark for the slowest decaying mode. Likewise, for state II we may assume a highly-delocalized state, such as an excitation uniformly distributed in several (but not all) sites of the lattice.} An example of ME is shown in Fig.2(a) for a lattice of size $L=9$ and for a dephasing rate $\gamma=J$. The figure depicts the temporal evolution of the distance $D$ for the two initial states 
  $\hat{\rho}_I(0)=|n_0 \rangle \langle n_0|$ with $n_0=(L+1)/2=5$ and $\hat{\rho}_{II}(0)=1/(L-1) \sum_{n=1}^{L-1} |n \rangle \langle n|$. 
     \begin{figure}[h]
 \centering
    \includegraphics[width=0.45\textwidth]{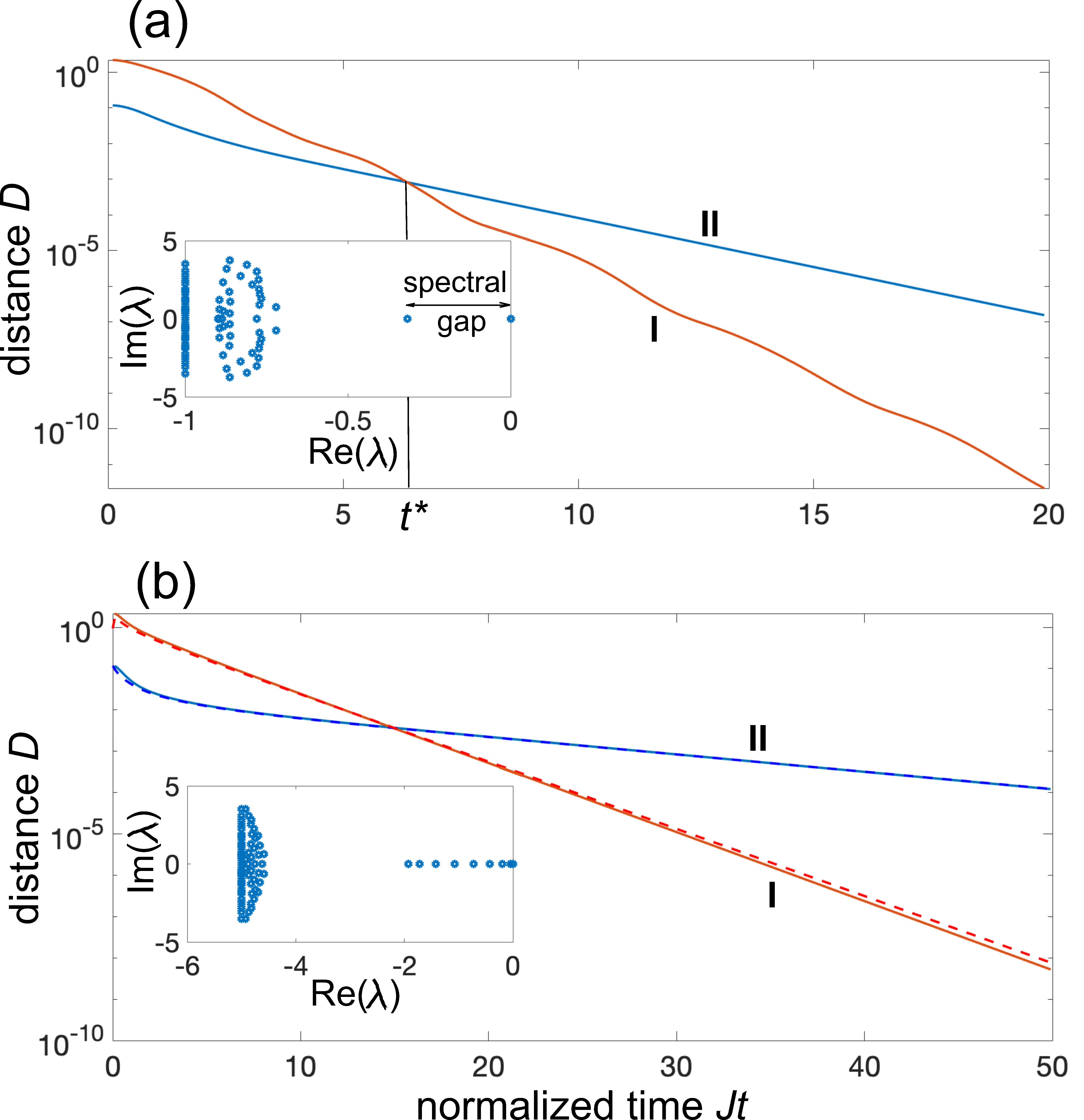}
   \caption{  \small (a) Numerically-computed behavior of the distance function $D_{\hat{\rho}}(t)$ [Eq.(4)] versus normalized time $Jt$ for the two initial states $\hat{\rho}_I(0)=|n_0 \rangle \langle n_0|$ with $n_0=5$ (curve I) and $\hat{\rho}_{II}(0)=1/(L-1) \sum_{n=1}^{L-1} |n \rangle \langle n|$ (cuvre 2). Lattice size $L=9$, dephasing rate $\gamma=J$. The inset shows the eigenvalues $\lambda$ of the Liouvillian $\mathcal{L}$ in complex plane. The spectral gap determines the relaxation rate of the slowest decaying mode. (b) Same as (a), but in the strong dephasing limit $\gamma=5J$, corresponding to a classical  random walk. The dashed curves in the main plot, almost overlapped with the solid ones, show the temporal evolution of the distance $D_{\hat{\rho}}$ as predicted by the classical  master equation. {\color{black} {A comparison of (a) and (b) shows that quantum coherence introduces some oscillatory features in the dynamics, however the ME persists in the limit of classical random walk.}}}
\end{figure}
  The eigenvalue spectrum of the Liouvillian $\mathcal{L}$ is shown in the inset of Fig.2(a). Note that the state $\hat{\rho}_I(0)$ is a highly-localized pure state with the photon initially prepared at site $n_0=(L+1)/2=5$, whereas  $\hat{\rho}_{II}(0)$ is a highly-delocalized mixed state very close to the stationary state $\hat{\rho}^{(S)}$.  Intuitively, one would expect the latter state to diffuse faster and fill the missing occupation of site $n=L$, however this is not the case. In fact, the state $\hat{\rho}_{I}(0)$, even if being a highly-localized state, spreads quikly in the lattice since it is not  overlapped with the slowing decaying mode  of the Liouvillian. Conversely, $\hat{\rho}_{II}(0)$ relaxes at a slower rate, which is established by the spectral gap of $\mathcal{L}$. Correspondingly, an intersection of the curves $D_{\hat{\rho}_I}$ and $D_{\hat{\rho}_{II}}$ at the time $t=t^* \simeq 6.36/J$ is observed, which is the clear signature of the ME.  {\color{black} We mention that the ME can be observed for larger system sizes $L$, however as $L$ increases the relaxation dynamics becomes slower and slower, making the crossing time $t^*$ rapidly increasing with $L$. Hence, for an experimental observation of the ME a small system size is more favorable. It should be also mentioned that the ME is observed using different measures for the distance $D_{\hat{\rho}} (t)$, such as the simple Hilbert-Schmidt distance \cite{S9}.}\  
  To understand the appearance of the ME, it is worth considering the strong dephasing limit $\gamma /J \gg 1$, where the photon hopping on the lattice becomes incoherent and reduces to a classical random walk. In such a regime, coherences $\rho_{n,m}(t)$ with $n \neq m$ are small and the evolution equations for the diagonal elements (photon occupation probabilities) $P_n(t) \equiv \rho_{n,n}(t)$ are described by the classical master equation (see for example \cite{S26})
  \begin{eqnarray}
  \frac{dP_n}{dt}  =  -2J_{inc}P_n+J_{inc} (P_{n+1}+P_{n-1}) \;\;\; (1<n<L) \nonumber \\
  \frac{dP_1}{dt}  =  -J_{inc}P_1+J_{inc} P_2 \; , \;   \frac{dP_L}{dt}  =  -J_{inc}P_L+J_{inc} P_{L-1}  \;\;\;\;
  \end{eqnarray}   
where $J_{inc}=2J^2/ \gamma$ is the incoherent hopping rate of the photon between adjacent sites of the lattice. Note that, since for $\gamma \gg J$ the density matrix is diagonal at leading order, the distance function$D_{\hat{\rho}}(t)$ takes the simple form
\begin{equation}
D_{\hat{\rho}}(t)= \log L+ \sum_{n=1}^{L} P_n(t) \log P_n(t).
\end{equation}
 Interestingly, in the strong dephasing limit  analytical expressions of the most relevant $L$ eigenmodes and eigenvalues of the Liouvillian can be obtained from the Markov transition matrix entering in Eq.(5), as shown in Sec.1 of the Supplemental document. The analysis shows that in a lattice with an odd number $L$ of sites the highly-localized initial state  $\hat{\rho}_I(0)=|n_0 \rangle \langle n_0|$, with $n_0=(L+1)/2$, does not excite the slowest decaying mode of the Liouvillian, thus the relaxation process (photon diffusion) toward the stationary state  is not limited by the spectral gap and is faster than the one of state  $\hat{\rho}_{II}(0)$. An example of the ME in the classical random walk limit is shown in Fig.2(b). Finally, we mention that, while previous analysis assumed  a single photon, the ME in the photonic random walk can be observed for any other (classical or non-classical) states of light, as shown in Sec.2 of the Supplemental document.\\
 \\
 {\it Mpemba effect in coupled-fiber loop mesh lattices}.  To illustrate the ME in an experimentally-accessible photonic setting, let us consider light dynamics in synthetic mesh lattices 
  based on optical pulse circulation in coupled fiber loops \cite{S28,S29,S30,S31,S32,S33,S34}. {\color{black}  As compared to other photonic systems, such as coupled waveguide lattices, it offers a few key advantages, such as a simple implementation of controllable dephasing via random phase modulation \cite{S19}, and the ability to maintain coherent pulse propagation over long propagation times \cite{S29,S30,S31,S32,S33,S34,S35}.}
   The system  consists of two fiber loops of slightly different lengths $L \pm \Delta L$, that are connected by a variable directional coupler with a coupling angle $\beta$, as schematically shown in Fig.3(a). A phase  modulator is placed in one of the two loops, which introduces stochastic phase changes thus realizing incoherent (dephasing) dynamics \cite{S19}.  When one or more pulses are injected into one loop, they will evolve into a pulse train after successive pulse splitting and interference at the variable coupler, realizing a synthetic lattice via time multiplexing \cite{S28,S29,S30,S31,S32,S33,S34}. Namely, the physical time is discretized as $t_n^{(m)}= mT +n\Delta T$, where $T= L/c$ is mean travel time and $\Delta T=\Delta L/c \ll T$ is the travel-time difference of light pulses in two loops. The pulse dynamics can thus be mapped into a "link-node" binary lattice model $(n, m)$, where $n$, $m$ denote the transverse lattice site and longitudinal evolution step (see e.g. \cite{S28,S29,S30,S31,S32,S33,S34}). Light dynamics is described by the set of discrete-time equations (see e.g. \cite{S26,S35})
 \begin{eqnarray}
u_n^{(m+1)} & = & \left( \cos \beta_{n+1} u_{n+1}^{(m)}+i \sin \beta_{n+1} w_{n}^{(m)} \right) \exp(i \phi_n^{(m)}) \;\;\;\;\;\; \\
w_n^{(m+1)} & = & i \sin \beta_{n} u_{n}^{(m)}+ \cos \beta_{n} w_{n-1}^{(m)}.
\end{eqnarray}
where $u_n^{(m)}$ and $w_n^{(m)}$ are the pulse amplitudes at discrete time step $m$ and lattice site (unit cell) $n$ in the two fiber loops, $\beta_n$ is the site-dependent coupling angle, and $\phi_n^{(m)}$ are uncorrelated stochastic phases with uniform distribution in the range $(-\pi, \pi)$.  A finite number $L$ of unit cells  in the lattice is simply obtained by assuming $\beta_n= \pi/2$ for $n \leq 1$ and $n \geq (L+1)$: in this case the site index $n$ can be restricted to  the range $n=1,2,...,L$.
 When the stochastic phases are applied at every time step, the incoherent dynamics is described by the map 
 \begin{eqnarray}
X_n^{(m+1)} & = &  \cos^2 \beta_{n+1} X_{n+1}^{(m)}+ \sin^2 \beta_{n+1} Y_{n}^{(m)} \\
Y_n^{(m+1)} & = & \sin^2 \beta_{n} X_{n}^{(m)}+ \cos^2 \beta_{n} Y_{n-1}^{(m)} 
\end{eqnarray}
  for the light pulse intensities $X_n^{(m)}=\overline{|u_n^{(m)}|^2}$ and $Y_n^{(m)}=\overline{|w_{n}^{(m)}|^2}$, where the overline denotes statistical average. 
  Assuming a uniform coupling angle $\beta_n$, close to $\pi/2$, for $n=2,3,...,L$, i.e. after letting
\begin{equation}
\beta_n= \pi/2-\theta \label{S21}
\end{equation}
with $| \theta|  \ll \pi/2$, the light pulse intensity distribution $P_n^{(m)}=X_n^{(m)}+Y_n^{(m)}$ in the various lattice unit cells varies slowly at each time step, so that we can treat $m=t$ as a continuous time variable \cite{S26}. After letting $P_n(t)=P_n^{(m)}$, the evolution equations for $P_n(t)$ can be approximated by the following set of differential equations (see Sec.3 of the Supplemental document)
  \begin{eqnarray}
  \frac{dP_1}{dt} & = & -\frac{1}{2} \theta^2 P_1 + \frac{1}{2} \theta^2 P_2 \nonumber \\
  \frac{dP_n}{dt} & = & - \theta^2 P_n + \frac{1}{2} \theta^2 (P_{n+1}+P_{n-1}) \; \; (1<n<L) \\ 
  \frac{dP_L}{dt} & = & -\frac{1}{2}  \theta^2 P_L +\frac{1}{2} \theta^{2} P_{L_1} \nonumber
  \end{eqnarray}  
       \begin{figure}[h]
 \centering
    \includegraphics[width=0.45\textwidth]{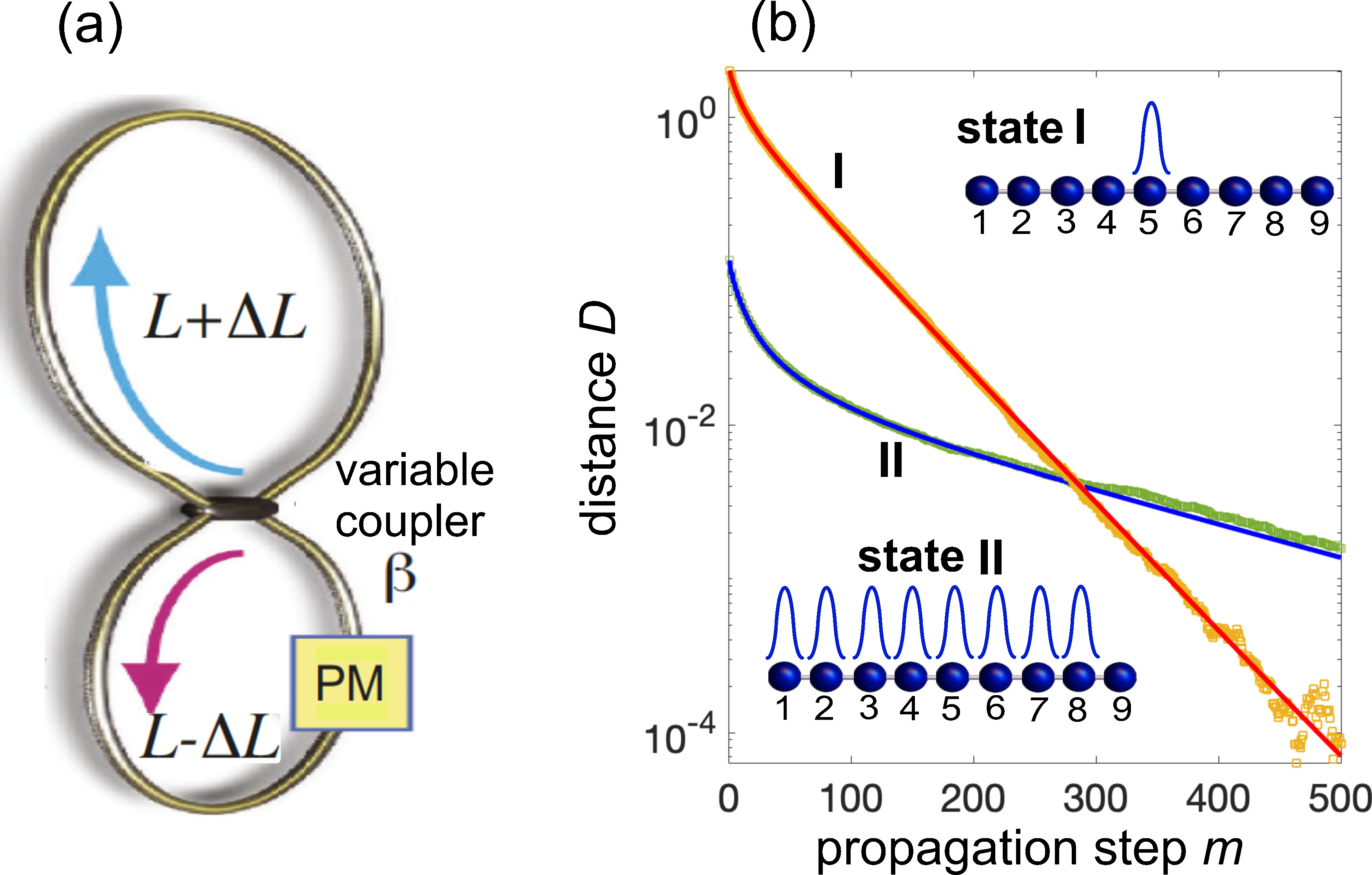}
   \caption{  \small (a) Schematic of two coupled fiber loops that realize photonic quantum walk with dephasing. Stochastic phases are applied by a phase modulator (PM) at each pulse crossing.  (b) Numerically-computed evolution of the distance function $D$ versus propagation time step $m$, as obtained using Eq.(6), for the two different initial conditions $u_n^{(0)}=\delta_{n,5}$, $w_n^{(0)}=0$ (curve I), corresponding to single pulse excitation at time slot $n_0=5$, and $u_n^{(0)}=(1/8) \sum_{l=1}^{8}\delta_{n,l}$,  $w_n^{(0)}=0$ (curve II), corresponding to initial excitation  with 8 equal pulses in each time slot of the lattice, except for the slot at $n=9$ where the pulse is missing (see insets). Lattice unit cell number is $L=9$, coupling angle $\beta= \pi/2-\theta$ with $\theta=0.2$. The solid curves in (b) are obtained by numerically solving the master equations (9) and (10), whereas the squares are obtained from numerical simulations of the stochastic equations (7) and (8) after averaging over 5000 different realizations of random phases.}
\end{figure}
which precisely correspond to the classical master equation (5) with an incoherent hopping rate $J_{inc}=\theta^2/2$. In an experiment, from the time-resolved measurements of pulse intensity distributions in the various sites of the lattice and after statistical average over stochastic realizations of the random phases $\phi_n^{(m)}$,  one can compute the distance function $D$ via Eq.(6) with 
$P_n(t)=\overline{|u_n^{(t)}|^2}+\overline{|w_{n}^{(t)}|^2}$ , thus revealing the ME for the incoherent hopping dynamics (diffusion) of photons toward the equilibrium state $P_n^{(S)}={\rm const}$. As an example, in Fig.3(b) it is shown the appearance of ME in a lattice comprising $L=9$ unit cells for a uniform coupling angle $\beta= \pi/2-\theta$ with $\theta=0.2$. The two initial states, schematically depicted in the insets of Fig.3(b), are: $u_n^{(0)}=\delta_{n,5}$, $w_n^{(0)}=0$ (state I), corresponding to single pulse excitation at time slot $n_0=5$ (the far-from-equilibrium state); and $u_n^{(0)}=(1/8) \sum_{l=1}^{8}\delta_{n,l}$,  $w_n^{(0)}=0$ (state II), corresponding to initial excitation with 8 pulses at time slots $n=1,2,..,8 $ (the close-to-equilibrium state). The distance function $D(t)$ versus propagation time step $t=m$ for the two evolving states has been computed by either numerical simulations of the stochastic equations (7) and (8), taking a statistical average of pulse intensities over 5000 realizations of stochastic phases, or by numerical simulations of the master equations (9) and (10). The numerical results show, as expected, excellent agreement between the two methods, indicating that ME should be feasible for an experimental observation.\\  
\\
{\em Conclusion.}
The Mpemba effect is a rather counterintuitive physical phenomenon in statistical physics for which a far-from-equilibrium state can
relax toward equilibrium faster than a state closer to equilibrium. This effect has been studied extensively in several classical and quantum systems, however it has been largely overlooked by the photonic community.  Here we showed that the Mpemba effect can be observed for light as well. Specifically, we considered the process of light diffusion in finite-sized photonic lattices under  incoherent (dephasing) dynamics, and showed that  certain highly-localized initial light distributions can spread and reach the equilibrium distribution faster than other delocalized distributions. 
{\color{black} {Owing to the ubiquity of dissipative dynamics in optics, it is envisaged that the ME could be observed in other dissipative photonic models, 
  suggesting new ways of light control such as  fast light transport in optical media,
 fast switching or reconfiguration in phase-change materials, and rapid thermalization of light in weakly-nonlinear multimode optical systems \cite{S36,S37}.}}\\ 
 \\
\noindent
{\bf Disclosures}. The author declares no conflicts of interest.\\
\\
{\bf Data availability}. No data were generated or analyzed in the presented research.\\
\\
{\bf Funding}. Agencia Estatal de Investigacion (MDM-2017-0711).\\
\\
{\bf Supplemental document}. See Supplement 1 for supporting content.

\newpage


 {\bf References with full titles}\\
 \\
 \noindent
 
1. E. B. Mpemba and D. G. Osborne, Cool?, Phys. Educ. {\bf 4}, 172
(1969).\\
2. G. S. Kell, The Freezing of Hot and Cold Water, Am. J. Phys. {\bf 37}, 564 (1969).\\
3. J. Bechhoefer, A. Kumar, and R. Ch\'etrite,  A fresh understanding of the Mpemba effect,
Nature Rev. Phys. {\bf 3}, 534 (2021).\\
4. A. Lasanta, F. Vega Reyes, A. Prados, and A. Santos,  
 When the Hotter Cools More Quickly: Mpemba Effect in Granular Fluids,
Phys. Rev. Lett. {\bf 119}, 148001 (2017).\\
5. Z. Lu and O. Raz, Nonequilibrium thermodynamics of
the markovian Mpemba effect and its inverse, Proc.
 Nat. Ac. Sci. PNAS {\bf 114}, 201701264
(2017).\\
6. A. Lapolla and A. Godec, Faster Uphill Relaxation in Thermodynamically Equidistant Temperature Quenches,
Phys. Rev. Lett. {\bf 125}, 110602 (2020).\\
7. A. Nava and M. Fabrizio, Lindblad dissipative dynamics
in the presence of phase coexistence, Phys. Rev. B
{\bf 100}, 125102 (2019).\\
8. S.K. Manikandan,
Equidistant quenches in few-level quantum systems, Phys. Rev. Res. {\bf 3}, 043108 (2021).\\
9. F. Carollo, A. Lasanta, and I. Lesanovsky, Exponentially
accelerated approach to stationarity in markovian open
quantum systems through the Mpemba effect, Phys. Rev.
Lett. {\bf 127}, 060401 (2021).\\
10. A.K. Chatterjee, S. Takada, and H. Hayakawa,
Quantum Mpemba Effect in a Quantum Dot with Reservoirs,
Phys. Rev. Lett. {\bf 131}, 080402 (2023).\\
11. F. Ares, S. Murciano, and P. Calabrese, 
Entanglement asymmetry as a probe of symmetry breaking, 
Nature Commun. {\bf 14},  2036 (2023).\\
12. L. Kh Joshi, J. Franke, A. Rath, F. Ares, S. Murciano, F. Kranzl, R. Blatt, P. Zoller, B. Vermersch, P. Calabrese, C.F. Roos, and M.K. Joshi, 
Observing the quantum Mpemba effect in quantum simulations, Phys. Rev. Lett. {\bf 133}, 010402, (2024).\\
13. F. Ivander, N. Anto-Sztrikacs, and D. Segal, 
Hyperacceleration of quantum thermalization dynamics by bypassing long-lived coherences: An analytical treatment,
Phys. Rev. E {\bf 108}, 014130 (2023).\\
14. J. Zhang, G. Xia, C.-W. Wu, T. Chen, Q. Zhang, Y. Xie, W.-B. Su, W. Wu, C.-W. Qiu, P. xing Chen, W. Li, H. Jing, and Y.-L. Zhou, 
Observation of quantum strong Mpemba effect, arXiv:2401.15951 (2024).\\
15. S.A. Shapira, Y. Shapira, J. Markov, G. Teza, N. Akerman, O. Raz, and R. Ozeri, The inverse Mpemba effect demonstrated on a single trapped ion qubit, Phys. Rev. Lett. {\bf 133}, 010403 (2024).\\
16. F. Ares, V. Vitale, and S. Murciano,
The quantum Mpemba effect in free-fermionic mixed states, arXiv:2405.08913 (2024).\\
17. M. Moroder, O. Culhane, K. Zawadzki, and J. Goold, The thermodynamics of the quantum Mpemba effect, 	arXiv:2403.16959 (2024).\\
18. M. A. Broome, A. Fedrizzi, B. P. Lanyon, I. Kassal, A. Aspuru-Guzik, and A. G. White, Discrete Single-Photon Quantum Walks with Tunable Decoherence,
Phys. Rev. Lett. {\bf 104}, 153602 (2010).\\
19.  A. Schreiber, K. N. Cassemiro, V. Potocek, A. Gabris, I. Jex, and Ch. Silberhorn, 
Decoherence and Disorder in Quantum Walks: From Ballistic Spread to Localization,
Phys. Rev. Lett. {\bf 106}, 180403 (2011).\\
20. J. Svozilik, R. de J. Leon-Montiel, and J.P. Torres, Implementation of a spatial two-dimensional quantum random walk with tunable decoherence, Phys. Rev. A {\bf 86}, 052327 (2012).\\
21. D.N. Biggerstaff, R. Heilmann, A.A. Zecevik, M. Gr\"afe, M.A. Broome, A. Fedrizzi, S. Nolte, A. Szameit, A.G. White, and I. Kassal, Enhancing coherent transport in a photonic network using controllable decoherence, Nature Commun. {\bf 7}, 11282 (2016).\\
22. F. Caruso, A. Crespi, A.G. Ciriolo, F. Sciarrino, and R. Osellame, Fast escape of a quantum walker from an integrated photonic maze,
Nature Commun. {\bf 7}, 11682 (2016).\\
23. S. Longhi, Incoherent non-Hermitian skin effect in photonic quantum walks, Light: Sci. \& Appl. {\bf 13},  95 (2024).\\ 
24. L. Fedichkin, D. Solenov, and C. Tamon, Mixing and decoherence in continuous time quantum walks on cycles, Quantum Inf. Comp. {\bf 6}, 263 (2006).\\
25. V. Kendon, Decoherence in quantum walks - a review, Math. Struct. Comp. Sci. {\bf 17}, 1169 (2007).\\
26. R. Zhang, H. Qin Hao, B. Tang and P. Xue,  Disorder and decoherence in coined quantum walks, Chin. Phys. B {\bf 22}, 110312 (2013).\\ 
27. S. Longhi, Dephasing-induced mobility edges in quasicrystals, Phys. Rev. Lett. {\bf 132}, 236301  (2024).\\
28. L.-N. Wu and A. Eckard,
Prethermal memory loss in interacting quantum systems coupled to thermal baths, Phys. Rev. B {\bf 101},  220302(R) (2020).\\
29.  A. Regensburger, C. Bersch, B. Hinrichs, G. Onishchukov, A. Schreiber, C. Silberhorn, and U. Peschel, Photon propagation in a discrete fiber network: an interplay of coherence and losses, Phys. Rev. Lett. {\bf 107}, 233902 (2011).\\
30. M. Wimmer, H.M. Price, I. Carusotto, and U. Peschel, Experimental measurement of the Berry curvature from anomalous transport,
Nature Phys. {\bf 13}, 545 (2017).\\
31. S. Weidemann, M. Kremer, T. Helbig, T. Hofmann, A. Stegmaier, M. Greiter, R. Thomale, and A. Szameit,
Topological funneling of light, Science {\bf 368}, 311 (2020).\\ 
32. S. Weidemann, M. Kremer, S. Longhi, and A. Szameit,
 Coexistence of dynamical delocalization and spectral localization through stochastic dissipation,
Nature Photon. {\bf 15}, 576 (2021).\\
33. S. Wang, C. Qin, W. Liu, B. Wang, F. Zhou, H. Ye, L. Zhao, J. Dong, X. Zhang, S. Longhi, and P. Lu, High-order dynamic localization and tunable temporal cloaking in ac-electric-field driven synthetic lattices, Nature Commun. {\bf 13}, 7653 (2022).\\
34. S. Weidemann, M. Kremer, S. Longhi, and A. Szameit, Topological triple phase transition in non-Hermitian Floquet quasicrystals,
Nature {\bf 601}, 354 (2022).\\
35. S. Wang, C. Qin, L. Zhao, H. Ye, S. Longhi, P. Lu, and B. Wang, Photonic Floquet Landau-Zener tunneling and temporal beam splitters,
Science Ad. {\bf 9}, eadh0415 (2013).\\ 
36. A.V. Pankov, I.D. Vatnik, D.V. Churkin, and S.A. Derevyanko,  Anderson localization in synthetic photonic lattice with random coupling,
Opt. Express {\bf 27}, 4424 (2019).\\
37. F.O. Wu, A.U. Hassan, and D.N. Christodoulides, Thermodynamic theory of highly multimoded nonlinear optical systems, Nature Photon. {\bf 13}, 776 (2019).\\
38. A. Ramos, L. Fern\'andez-Alc\'azar, T. Kottos, and B. Shapiro, Optical Phase Transitions in Photonic Networks: a Spin-System Formulation,
Phys. Rev. X {\bf 10}, 031024 (2020).\\


\end{document}